\documentclass[pre,aps,amsmath,amssymb,onecolumn]{revtex4}

\def\b{\mathbf}
\def\e{\epsilon}

\usepackage{graphicx}
\usepackage{dcolumn}
\usepackage{bm}
\usepackage{color}
\usepackage{epsfig,graphics,amssymb,amsmath,subeqnarray,setspace,graphicx,amsthm,epstopdf,subfigure}

\begin{document}

\title{Elastocapillary self-folding: buckling, wrinkling, and collapse of floating filaments}

\author{Arthur A. Evans$^1$}
\email{aevans@physics.ucsd.edu}
\author{Saverio E. Spagnolie$^2$}
\email{spagnolie@math.wisc.edu}
\author{Denis Bartolo$^3$}
\author{Eric Lauga$^4$}
\affiliation{$^1$ Department of Chemistry and Biochemistry, University of California Los Angeles, 405 Hilgard Ave, Los Angeles, CA 90095, USA.}
\affiliation{$^2$ Department of Mathematics, University of Wisconsin-Madison, 480 Lincoln Drive, Madison, WI 53706, USA.}
\affiliation{$^3$ PMMH, CNRS, ESPCI ParisTech, Universit\'es Paris 6 \& Paris 7, 10 rue Vauquelin, 75231 Paris cedex 05, France.}
\affiliation{$^4$ Department of Mechanical and Aerospace Engineering, University of California San Diego, 9500 Gilman Drive, La Jolla, CA 92093, USA.}

\date{\today}

\begin{abstract}

When a flexible filament is confined to a fluid interface, the balance between capillary attraction,  bending resistance, and tension from an external source can lead to a self-buckling instability. We perform an analysis of this instability and provide analytical formulae that compare favorably with the results of detailed numerical computations. The stability and long-time dynamics of the filament are governed by a single dimensionless elastocapillary number quantifying the ratio between capillary to bending stresses. Complex, folded filament configurations such as loops, needles, and racquet shapes may be reached at longer times, and long filaments can undergo a cascade of self-folding events.

\end{abstract}

\maketitle


\section{Introduction}

A common strategy for assembling intricate structures at the micro- and nano-scale is to exploit self-assembly. By introducing colloidal building blocks into soft media such as fluid interfaces, nematic liquid crystals, or more complex mesophases \cite{Whitesides2002,DietrichColloids,SurfactantColloidCrystals,CapillaryInteractionsInclusions,Lapointe2009,ColloidalDispersions}, the resulting distortions of the medium can be used to fabricate more elaborate colloidal objects. Over the last ten years much effort has been devoted to achieving complex self-assembly by tailoring the shape of the elementary colloidal building blocks \cite{Lapointe2009,LockKeyColloids}. Alternatively, the existence of mechanical instabilities in elastic materials suggests a simple but elegant method for guiding the creation of complex objects: a bottom-up assembly driven by elastic instabilities in simple, flexible structures. Such a self-assembly might begin with simple building block configurations, such as straight filaments or flat sheets, which when coupled to another medium would fold or wrinkle to minimize the total energy of the system. If the system is well characterized theoretically, the dominant force balance can be tuned to yield desirable shapes. Recent examples include the use of adhesion and delamination \cite{WrinkleFoldTransition,LocalizedBuckling,CapillaryWrinklingMembranes,CompressionIntegrableDaimantWitten,PocivavsekFolding,BrauParametric,EvansLauga2009}, and swelling and capillary interactions \cite{ElastoCapReview,MeniscusLithography,NanopillarAssembly}.

When particles are confined to a fluid interface, their interactions are mediated through the surface deformations \cite{DietrichColloids,VellaMahadevan}. The deformations and thus the surface energy of the fluid may be decreased by a rearrangement of the particles on the surface, leading to effective capillary forces which can be attractive or repulsive depending on the geometry, density differences, and wetting properties of the particles. For two identical particles, the forces are always attractive.  This so-called ``Cheerios effect,'' and similar surface-mediated aggregation have been investigated in the context of vesiculation \cite{MembraneColloidAggregation,MembraneVesiculation}, colloidal flocculation \cite{DietrichColloids,ColloidalDispersions}, and millimetric ecology \cite{MeniscusClimb,Voise2011}. 
 While there have been many studies on filaments and polymers that interact via short-range forces and/or external flow fields~\cite{MackintoshI,MackintoshII,TensionStraighteningNelson,TensionPropagationKroy,TensionDynamicsKroy,YoungFilamentInteractions,BauschKroyNatPhys,Julicher2007,KantslerGoldsteinFilaments}, the impact of long-range capillary forces on the shapes of  flexible filaments lying on a fluid interface remains unexplored.

In this paper we consider the problem of elastocapillary self-folding of a flexible micro-filament.  
The model system, illustrated in Fig.~\ref{Figure1}, is that of a flexible filament confined to a fluid interface. The balance between {long-range} capillary forces, which are attractive between  different parts of the filament, and  bending stresses resisting deformation can lead to a novel buckling instability (as illustrated in Fig.~\ref{Figure1}b), and at longer times to highly deformed folded patterns in the filament.  Herein we show how to harness the shape instability to fabricate folded structures.
The paper is organized as follows. In \S II we introduce the setup and discuss the energetics of the system. In \S III we pose the non-dimensional problem to solve for the shape dynamics in a viscous fluid. The linear stability analysis for the shape is presented in \S IV, where we determine the functional dependence of the most unstable wavelength on both filament and fluid properties, as well as the effect of external tensile forces. We show that the threshold for self-buckling depends on a single dimensionless parameter quantifying the ratio between self-attraction and bending akin to the recently introduced elastocapillary number~\cite{ElastoCapReview}. The complex, folded filament configurations reached far from the threshold of instability are explored by numerical simulations in \S V. We close with a discussion of our main findings in \S VI.

\section{Energy and scaling}

Consider two identical particles of size $a$ located at the interface between two fluids, for example air and water. Provided the distance between the particles, $R$, is much smaller than the capillary length, their effective capillary interactions are described by the  interaction energy
$E_{\rm int} = \gamma a^2\ln ({R}/{\ell_c})$, 
where $\gamma$ is the interaction strength per unit area and the capillary length is defined by $\ell_c=\sqrt{\sigma/\rho g}$, with $\sigma$ the surface tension of the interface, $\rho$ the difference in density between the two fluids, and $g$ the acceleration due to gravity ($\ell_c\approx 2$ mm for an air-water interface) \cite{deGennes}.
 The strength of the interaction comes from the competition between surface forces and an external force monopole \cite{DietrichColloids}. This force can originate from gravity, as in the case of the ``Cheerios effect'' \cite{VellaMahadevan}, or from electro- or magneto-static forces  \cite{CapillaryInteractionsInclusions}; note that $\gamma$  contains additional information such as the particle geometry and wetting properties.

\begin{figure}[t]
\includegraphics[width=.7\textwidth]{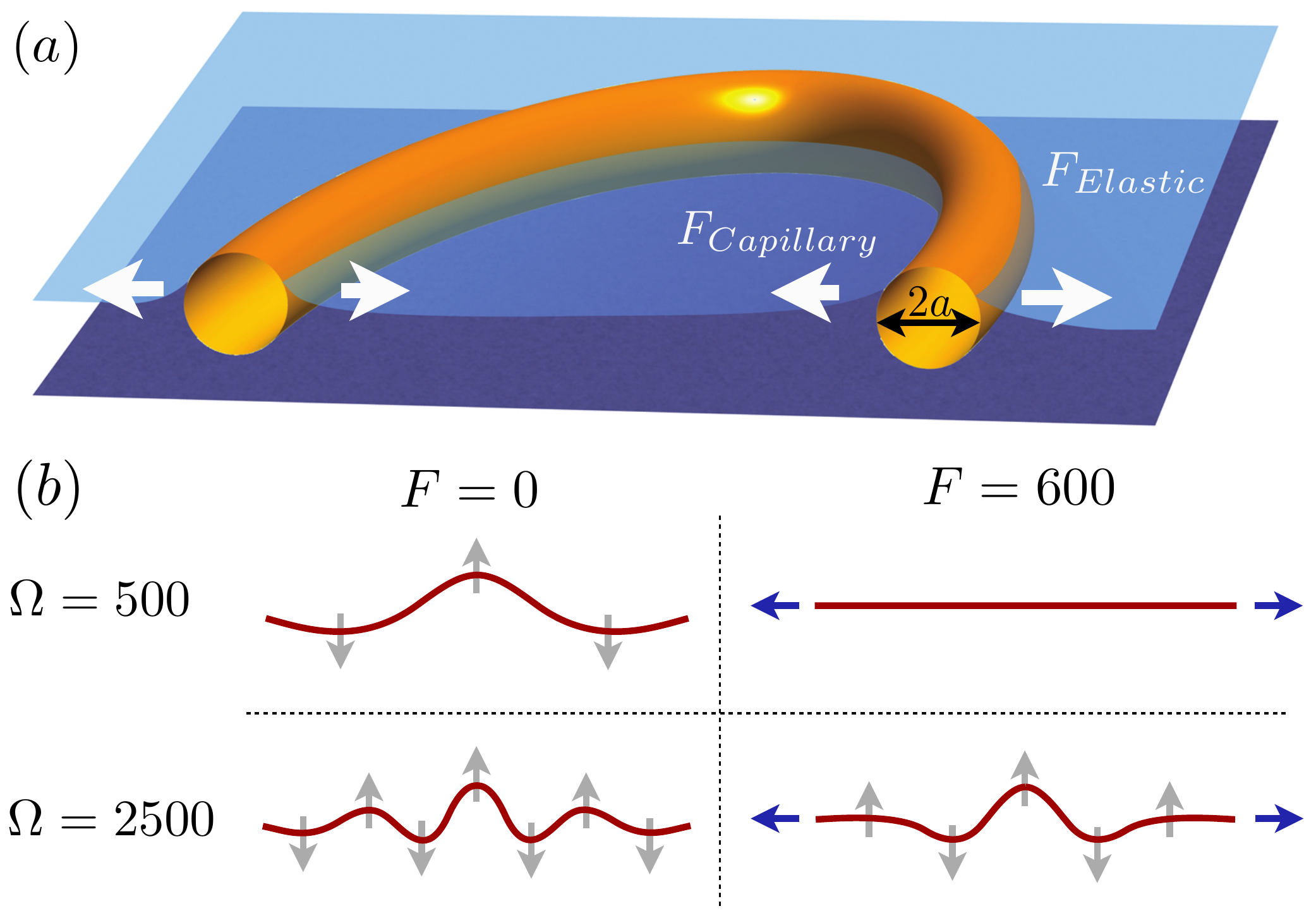}
\caption{The model system considered in the paper. (a): 
A flexible filament (of radius $a$) floating at the interface between two fluids. Every part of the filament experiences capillary attraction to every other part, while bending resists deformation. The competition between capillary forces and bending results in buckling and folding. 
(b): The most unstable wavelength depends on a dimensionless strength of self-attraction, $\Omega$. The onset of this instability can be tuned with the introduction of an added tension from an external source, $F$, which also decreases the most unstable wavelength (the shapes shown are obtained using the numerical computations).}
\label{Figure1}
\end{figure}

The model system considered in this paper is illustrated in 
Fig.~\ref{Figure1}. We consider  an inextensible elastic filament of length $L < \ell_c$ and bending modulus $B$. The filament bending energy is
\begin{gather}
E_{\rm bend}=\frac{B}{2}\int_0^L{|\mathbf{x}_{ss}|^2}ds+\int_0^L{T(s)[|\mathbf{x}_s|^2-1]}ds,
\end{gather}
where $\mathbf{x}(s,t)$ denotes the two-dimensional position vector of the filament on the fluid surface at time $t$. The arc-length coordinate parameterizing the filament is denoted by $s$, subscripts denote partial derivatives, and $T(s)$ is a Lagrange multiplier enforcing an inextensibility condition. For notational simplicity in the following we will commonly assume an implicit dependence upon time.
The contribution to the energy due to capillary interactions may be written as
\begin{gather}
E_{\rm int}=\frac{1}{2}\int_0^L{\int'{\gamma \ln\left[\frac{R(s,s')}{\ell_c}\right]ds'}ds},
\end{gather} 
where $R(s,s')=|\mathbf{x}(s)-\mathbf{x}(s')|$, $\gamma$ is now the energy per area of the interaction between two points on the filament, and where we have defined $\int'=\int_{-a}^{s-a}+\int_{s+a}^{L+a}$, with $a$ the size of the short-length cutoff ($a$ is typically of the order of the filament radius). 

Scaling distances by the filament length,  $L$, the tension $T(s)$ by $B/L^2$, and energies by $B/L$, the total energy combining  bending and interaction takes the simple form:
\begin{equation}\label{Energy}
E=\int_0^1\left\{\frac{1}{2}|\mathbf{x}_{ss}|^2+T(s)[| \mathbf{x}_s|^2-1]+\frac{\Omega}{2}\int' \ln R(s,s')ds'\right\}\,ds
.\end{equation}
$E$ depends on a single dimensionless parameter, $\Omega=\gamma L^3/B$, which quantifies the balance between capillarity self-attraction and bending, and is akin to the so-called elastocapillary number~\cite{ElastoCapReview}.

Next, we focus on deriving the equation of motion for a filament relaxing to equilibrium, which we will then exploit to analyze shape instabilities.

\section{Equations of motion}

To perform detailed calculations of the filament dynamics, we consider variations of the total energy, $E$, with respect to the centerline position vector, $\mathbf{x}(s)$, and tension, $T(s)$. The forces so derived must balance with the forces  generated by the motion through the fluid, which we model through a local drag coefficient, $\zeta \mathbf{x}_t$, an appropriate approach for dynamics in a very viscous fluid. The ambient fluid is assumed to be quiescent and of infinite extent. Scaling time upon the characteristic time scale $\tau= L \zeta/B$, we arrive at a dimensionless equation for the motion of the filament, 
\begin{gather}
 \mathbf{x}_t=-\mathbf{x}_{ssss}+\Omega\int'{\frac{\mathbf{x}(s')-\mathbf{x}(s)}{R(s',s)^2}ds'}+\partial_s(T(s)\mathbf{x}_s).\label{eom}
\end{gather}
Anticipating our numerical approach, the local part of the integration in Eq.~\ref{eom} may be handled analytically by adding and subtracting the singular part of the integrand. In so doing, we find
\begin{gather}
\int'  \frac{\b{x}(s')-\b{x}(s)}{R(s',s)^2}\,ds'=\b{K}[\b{x}(s')](s)+c(s)\b{x}_s(s)+O(\delta\log\delta),\\
\b{K}[\b{x}(s')](s)=\int_{-\delta}^{1+\delta}\Big\{ \frac{\b{x}(s')-\b{x}(s)}{R(s',s)^2}-\frac{\b{x_s}(s)}{s'-s}\Big\}\,ds',
\end{gather}
where $c(s)\equiv\log[(1+\delta-s)/(s+\delta)]$, with $\delta=a/L$ the dimensionless short wavelength cut-off. Rewriting the dimensionless filament position equation, we therefore have
\begin{gather}
 \b{x}_t=-\b{x}_{ssss}+\Omega c(s)\b{x}_s(s)+\Omega \b{K}[\b{x}(s')](s)+\partial_s\left(T(s)\b{x}_s\right).\label{filpos}
\end{gather}
Classically, the equation for the tension $T(s)$ is derived using the relation $\b{x}_s\cdot \b{x}_{ts}=0$, and also the identities
\begin{gather}
\b{x}_s\cdot \b{x}_{ss}=0,\,\,\,\,\b{x}_s\cdot \b{x}_{sss}=-\b{x}_{ss}\cdot\b{x}_{ss}, \\
\b{x}_{s}\cdot \b{x}_{ssss}=-3\b{x}_{ss}\cdot \b{x}_{sss},\,\,\,\,\b{x}_{s}\cdot \b{x}_{sssss}=-4\b{x}_{ss}\cdot \b{x}_{ssss}-3\b{x}_{sss}\cdot \b{x}_{sss},
\end{gather}
yielding the differential equation
\begin{equation}
\left(\partial_{ss}-|\b{x}_{ss}|^2\right)T(s)=-\left[4\b{x}_{ss}\cdot \b{x}_{ssss}+3|\b{x}_{sss}|^2\right]
-\Omega c_s(s)-\Omega \b{x}_s\cdot \frac{\partial}{\partial s} \b{K}[\b{x}(s')](s). \label{filten}
\end{equation}
Boundary conditions are derived by including terms to the energy functional to account for the work done in forcing the endpoints to move, namely, $E_F=-\b{F}_0 \cdot \b{x}(0)-\b{F}_1 \cdot \b{x}(1)$, where $\b{F}_0$ and $\b{F}_1$ are (dimensionless) externally-applied forces at the $s=0$ and $s=1$ endpoints of the filament, respectively. Assuming that the filament ends are free, hinged about a specified (possibly moving) point, or pulled by the forces defined above, the solvability conditions thus require
\begin{gather}
\b{x}_{ss}(0)=\b{x}_{ss}(1)=0,\label{bcs0}\\
-\b{F}_0+\b{x}_{sss}(0)-T(0)\b{x}_s(0)=0,\\
-\b{F}_1-\b{x}_{sss}(1)+T(1)\b{x}_s(1)=0.
\end{gather}
The boundary conditions are then found to be
\begin{gather}
\b{x}_{sss}(0)=\left(\b{I}-\b{x}_s(0)\b{x}_s(0)^T\right)\b{F}_0,\label{bcs2}\\
\b{x}_{sss}(1)=-\left(\b{I}-\b{x}_s(1)\b{x}_s(1)^T\right)\b{F}_1,\label{bcs3}\\
T(0)=-\b{F}_0\cdot \b{x}_s(0),\,\,\,\,\,T(1)=\b{F}_1\cdot \b{x}_s(1).\label{bcs1}
\end{gather}

To summarize, the partial differential equation describing the motion of a floating filament with either free or forced ends is given by Eq.~\ref{filpos} with the boundary conditions shown in Eqs.~\ref{bcs0},~\ref{bcs2}, and ~\ref{bcs3}. It is coupled to the equation for the tension, Eq.~\ref{filten}, with  boundary conditions shown in Eq.~\ref{bcs1}. The numerical method developed to solve this system of equations is described in Appendix A. In the next section we exploit our equations of motion to describe the linear stability of the filament.

\section{Linear stability analysis}

\subsection{Numerical investigation of growth rates}

\begin{figure*}[t]
\includegraphics[width=.9\textwidth]{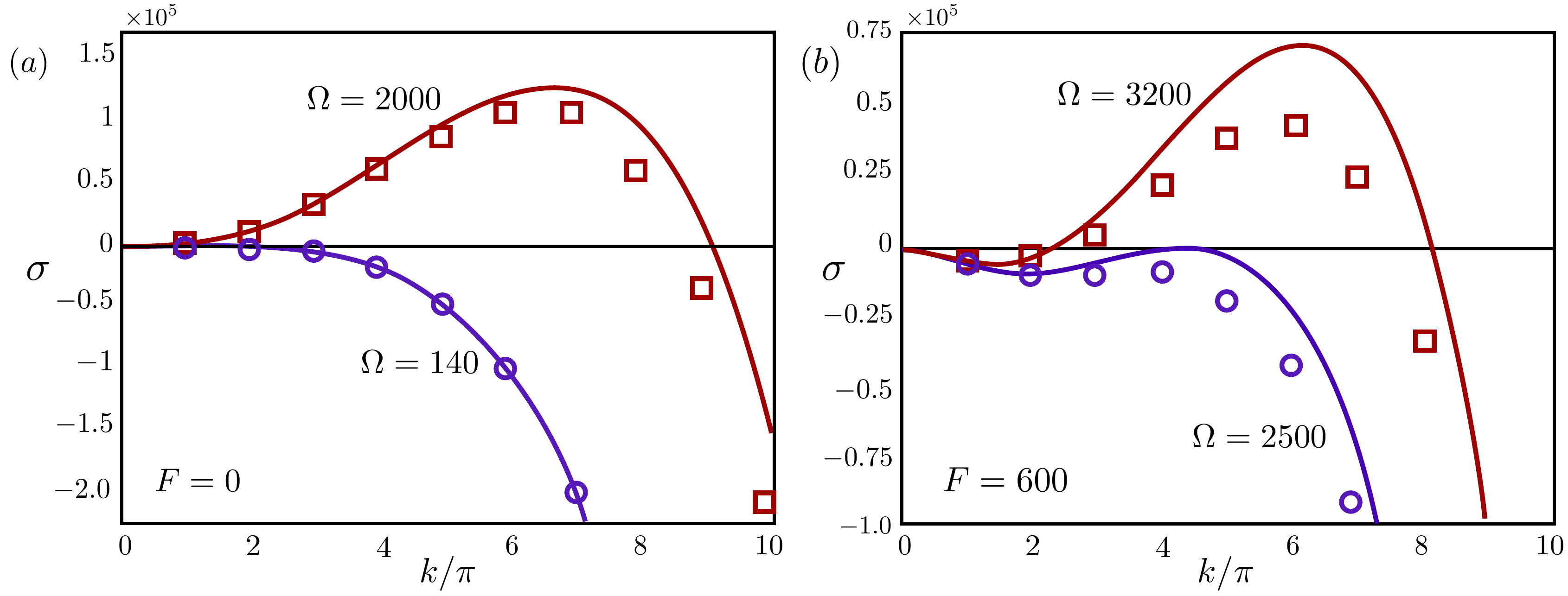}
\caption{Instability of the filament shape: Exponential growth rate, $\sigma$, computed  numerically  (symbols) and predicted analytically (Eq.~\ref{dispersion},  lines). (a): With no external force, $F=0$, we consider two interaction strengths $\Omega=140$ (just below the critical value $\Omega_c=144$ for the instability), and $\Omega=2000$. (b): With an external force $F=600$, we consider the values $\Omega=2500$ and $\Omega=3200$.}
\label{Figure2}
\end{figure*}

Before we pursue an analytical description of filament instability, let us begin with a brief numerical investigation using the framework described in Appendix A. In order to consider the growth rate of a given wavenumber $k$, we seed a nearly straight filament with the perturbed dimensionless shape $\b{x}(s)=(x(s),y(s))=(s,a_0\sin(k s))$, with $a_0=10^{-4}$. Ten timesteps are then taken with a step-size $\Delta t=10^{-9}$, and the growth rate is computed as $\sigma = \log(a_{10}/a_0)/(10\Delta t)$, where $a_{10}=2\int_0^1\sin(k s)y(s,10 \Delta t)\,ds$. Figure~\ref{Figure2}a shows (as symbols) the growth rates computed for two different interaction strengths when the filament ends are left free. For $\Omega=140$, we find that there are no wavenumbers for which the growth rate is positive, indicating that the filament is linearly stable to transverse perturbations. Despite the attraction of every part of the filament to every other part, the filament's stiffness prevents the buckling and subsequent collapse when the body is perturbed from a straight conformation. Conversely, increasing the interaction strength to $\Omega=2000$, the capillary attraction overcomes the bending elasticity and we observe that the filament is linearly unstable. A wide range of wavenumbers corresponds to positive growth rates. 
It is worth noting that the the critical value of the interaction strength above which capillary interactions destabilize the filament is much larger than 1, suggesting that it could not have been inferred  solely from  dimensional analysis. We also show in Fig.~\ref{Figure2}b the growth rates computed for two different interaction strengths, where an external pulling force has been introduced at the filament ends. For a given interaction strength the external force reduces the growth rate of every mode, and so decreases the value of the largest unstable mode. Meanwhile, setting $F>0$ introduces a negative slope in $\sigma$ for small $k$, and significantly changes the value of the smallest unstable mode. 

\subsection{Analytical predictions}
Having gained intuition from the numerical simulations described above, we now turn to an analytical investigation of filament instability. Our goal is to establish the dispersion relation governing the bending excitations of a floating elastic filament. 

One exact solution to the complete floating filament system is the possibly unstable trivial solution, that of a straight filament lying along a director $\b{e}$, namely
\begin{gather}
\b{x}^*(s,t)=s\,\b{e}-\b{e}\cdot[\b{F}_1+\b{F}_0]\,t,
\end{gather}
and
\begin{gather}
T^*(s)=\b{e}\cdot\left[(s\,\b{F}_1-(1-s)\b{F}_0\right]+\Omega\left\{ (1-s)\, c(s)+\log\left(\frac{s+\delta}{1+\delta}\right)+\delta\log\left[\frac{(1-s+\delta)(s+\delta)}{\delta(1+\delta)}\right]\right\}.\label{T0}
\end{gather}
From here, we are well-situated to perform a classical stability analysis based on perturbations of both $\b{x}(s)$ and $T(s)$. We assume that the endpoints are pulled upon by an external force of equal magnitude $F$, $\b{F}_0=-F\b{e}$, $\b{F}_1=F \b{e}$, and we set $\b{e}=\b{\hat{x}}$ for convenience of notation. 

To begin, we write $\mathbf{x}(s,t)=s\b{\hat{x}}+\epsilon \tilde{y}(s,t)\b{\hat{y}}+O(\epsilon^2)$, and let $T=T^*+\tilde{T}$, with $T^*$ the equilibrium tension from Eq.~\ref{T0}. By tracking terms of size $\e$ and smaller, it may be shown that at leading order, the  non-local contribution to the capillary force only gives rise to a transverse force contribution as
\begin{gather}
\b{K}[\b{x}(s')](s)=O(\e^2)\b{\hat{x}}+\e\int_{0}^{1}\left[\frac{\tilde{y}(s')-\tilde{y}(s)}{(s'-s)^2}-\frac{\tilde{y}_s}{s'-s}\right]\,ds'\b{\hat{y}}.\label{Kint}
\end{gather}
Upon the insertion of $\frac{\partial}{\partial s} \b{K}[\b{x}(s')](s)=O(\e^2)\b{\hat{x}}+O(\e)\b{\hat{y}}$ into Eq.~\ref{filten}, we deduce that $T=T^*+O(\e^2)$, such that we may dispose of tension perturbations $\tilde{T}$ for the linear stability analysis. Similarly, we find that $\hat{\b{x}}\cdot \b{x}_t=O(\e^2)$ from Eq.~\ref{filpos}, so that the linear stability consideration may focus solely on transverse filament motions through the evolution of $\tilde{y}(s,t)$.

In order to gain physical insight to the the nonlocal operator ${\bf K}$ given by Eq.~\ref{Kint}, we assume that $s$ is sufficiently far removed from the filament endpoints and look at its effect on plane waves of wavenumber $k$. By doing so, we find that the transverse component of $\b{K}[\b{x}(s')](s)$ may be written as
\begin{gather}
\b{K}[\b{x}(s')](s)\cdot{\hat{\bf y}}\approx g(k) \,\tilde{y}(s),
\end{gather}
where
\begin{gather} 
g(k) = 2\left[1-\cos(k)+\gamma + \log(k)-k \,S_i(k)-C_i(k)\right],
\end{gather}
and with the sine and cosine integrals defined by
\begin{gather}
S_i(k)=\int_0^k \frac{\sin(t)}{t}\,dt,\,\,\,\,\,C_i(k)=-\int_k^\infty \frac{\cos(t)}{t}\,dt. \label{sici}
\end{gather}
The mathematical details of this derivation are provided  in appendix B.
Combining the results above, the transverse component of the filament position equation, Eq.~\ref{filpos}, gives a linear relation for $\tilde{y}(s)$,
\begin{gather}
\tilde{y}_t=-\tilde{y}_{ssss}+ \Omega\,g(k)\,  \tilde{y}+\left[\Omega c(s)+T^*_s\right]  \tilde{y}_s+T^* \tilde{y}_{ss}+O(\e^2).
\end{gather}
Upon inspection of $T^*$, we are reminded that the local self-attraction term $c(s)$ could just as well have been absorbed into the tension, and simplifying the above results in 
\begin{gather}
\tilde{y}_t=-\tilde{y}_{ssss}+T^* \tilde{y}_{ss}+\Omega\,g(k)\,\tilde{y}.\label{linearized}
\end{gather}
We are thus left with the relatively simple over-damped Euler-Bernoulli beam equation, with a non-uniform tension, and with an additional term of magnitude $\Omega$ originating from the non-local capillary attraction. In the following we discuss in detail the impact of this novel effective elasticity term on the dynamic response of the floating filament.

To explore the response of the filament shape to small perturbations, we explicitly insert the ansatz $\tilde{y}= e^{i k s+\sigma t}$. While the finite-size effects must come into play for small wavenumbers and for longer times (trigonometric functions do not satisfy the boundary conditions on Eq.~\ref{linearized}, for example), this ansatz is expected to be  accurate far from the end points of the filament. Upon insertion of the equilibrium tension $T^*(s)$, an inner product of Eq.~\ref{linearized} against $e^{-i k s}$ reveals the dispersion relation for the growth rate,
\begin{gather}
\sigma=-k^4+\frac{1}{2}k^2\left(\Omega-2F\right)+\Omega\,g(k).\label{dispersion}
\end{gather}

How well does the analytical expression above capture the  results of the full numerical simulation? Returning to Figs.~\ref{Figure2}a and~\ref{Figure2}b, the analytically predicted growth rates from Eq.~\ref{dispersion} are shown as lines for each of the four cases considered. While discrepancies are expected due to end effects, which are not captured in our asymptotic consideration, in general we find good agreement between the numerical and analytical results. Therefore, we can use Eq.~\ref{dispersion}  with confidence to  demonstrates the fundamental physics behind the elastocapillary instability, which we do below.

\subsection{Discussion of the elastocapillary instability}

The quartic term and quadratic terms in Eq.~\ref{dispersion}  are associated with the canonical Euler buckling. However, as expected, the capillary attraction leads to an additional compressive load (the $\Omega k^2$ term in Eq.~\ref{dispersion}). Any attractive short-range coupling  along the filament is expected to renormalize the tension term in a similar fashion. We now discuss the consequence of the more complex contribution: $\Omega g(k)$,  to the dispersion relation. We first stress that $g(0)=0$, and $g(k)<0$ for all $k>0$, which means that  the non-local part of the capillary coupling acts to stabilize straight conformations.

\begin{figure*}[t]
\includegraphics[width=.9\textwidth]{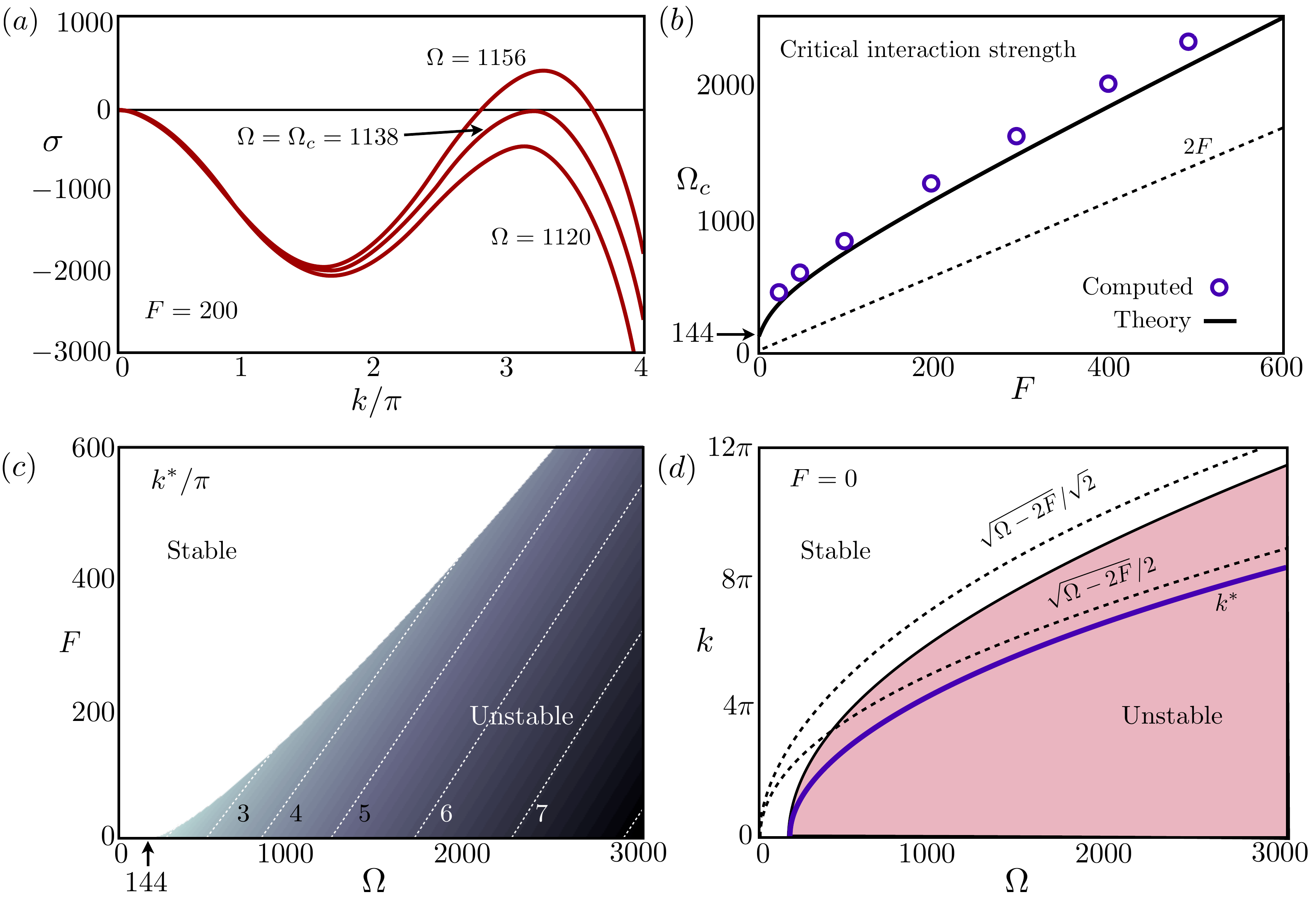}
\caption{(a): Growth rate of the instability, $\sigma$, as a function of wavenumber, $k$, for three values of the interaction strength, $\Omega$, with external force $F=200$ (the value of the critical interaction strength is $\Omega_c=1138$). (b):  Critical interaction strength, $\Omega_c$, below which there are no unstable modes as a function of the external force, $F$, from both numerical simulations (symbols) and analytical prediction (line). (c): Map of the most unstable mode, $k^*/\pi$, analytically predicted (from Eq.~\ref{dispersion}) as a function of $\Omega$ and $F$. (d) Setting $F=0$ (free filament ends), the most unstable wavenumber, $k^*$, is surrounded by an increasingly wide band of unstable modes as $\Omega$  increases (shaded region); analytical estimates for the most unstable and the highest unstable mode are shown as dashed lines.}
\label{Figure3}
\end{figure*}

In addition, given that $\Omega \geq 0$ and $F\geq 0$, the predicted growth rate $\sigma$ goes to zero as $k\rightarrow 0$, to $-\infty$ as $k\rightarrow \infty$, and may or may not take on positive values for finite $k$, depending on the values of $\Omega$ and $F$. For example, Fig.~\ref{Figure3}a shows the growth rate from Eq.~\ref{dispersion} as a function of wavenumber $k$ for three values of the interaction strength, $\Omega$, with an external pulling force $F=200$. For $\Omega$ less than a critical value, $\Omega_c=1138$ in this case, all modes are predicted to decay in time, and the filament is predicted to be stable. For $\Omega>\Omega_c$, we observe the appearance of a band of unstable modes and the filament becomes unstable. The critical interaction strength for instability, $\Omega_c$, is shown as a function of the value of $F$ in Fig.~\ref{Figure3}b. It takes the value $\Omega_c=144$ when $F=0$ and increases with increasing $F$. Both the values predicted by the analytical theory and those found by the full numerical simulation are shown; here again we see  that the analysis is sufficient to capture the linear instability of the filament dynamics. 

To develop physical insight into this behavior, we look at at the asymptotic form  of $\sigma(k)$ in the limit of long wavelengths. We find that  $\sigma\sim -F k^2+(\Omega/144-1)k^4$ for $k \ll 1$. In this regime, the capillary-induced compressive load vanishes; the self-attraction does not renormalize the tension of the filament in the long wavelength limit. Nevertheless, the capillary coupling softens the filament by effectively reducing its bending stiffness: the pre-factor of the quartic term decreases linearly with  $\Omega$, and changes its sign at $\Omega_c=144$. Above this critical value the bending excitations along the filament are amplified. As noted in the previous section, this unexpectedly high value of $\Omega_c$ could not have been anticipated on the basis of dimensional arguments alone. Furthermore, as $\partial \sigma/\partial k<0$ as $k\rightarrow 0$ for $F>0$ the smallest unstable mode, if there is one, must take on a finite value when an external pulling force is applied.

Beyond this prediction of the destabilization threshold, we can also accurately predict the value of the most unstable mode, which chiefly sets the ultimate folded shape of the filament as we will show more precisely in the next section. For a given force $F$ and interaction strength $\Omega$ (such that $\Omega>\Omega_c$), the most unstable mode $k^*$ is recovered upon setting $\partial \sigma/\partial k$ to zero and solving for k. The resulting value is presented in Fig.~\ref{Figure3}c as a function of both $\Omega$ and $F$.  Predictably, the most unstable mode increases with increasing self-attraction (increasing $\Omega$), and decreases as the filament ends are pulled upon with greater intensity (increasing $F$). The introduction of an external force renders the filament stable for small interaction strengths $\Omega$, and so intuitively prolongs the onset of buckling. Increasing $\Omega$ and/or decreasing $F$ is therefore expected to yield a more finely folded filament shape once the instability fully develops. 

In the specific case of a free filament, $F=0$, Fig.~\ref{Figure3}d shows the value of most unstable wavenumber as a function of $\Omega$.  The shaded region indicates the band of unstable modes for a given interaction strength $\Omega$. Again, these variations can be qualitatively understood by inspecting the asymptotic behavior of $\sigma(k)$. Indeed, for $k\gg1$ we find $\sigma(k) \sim -k^4+\Omega\left[k^2-\pi k+2(\log(k)+\gamma+1)\right]$, with $\gamma$ the Euler-Mascheroni constant. This latter approximation of $\sigma(k)$ is accurate to within $2\%$ of the exact value for wavenumbers as small as $k=\pi$, and it can therefore replace $\sigma(k)$ generally for all but the very longest perturbation wavelengths. For $k \gg 1 $,  the problem boils down to the canonical Euler Buckling induced by the longitudinal self-attraction. Hence, balancing the cubic and linear terms in the equation $\partial \sigma/\partial k=0$ gives the usual  approximation to the most unstable mode \cite{Love92}:
\begin{gather}
k^*\approx \frac{1}{2}\sqrt{\Omega-2F}.\label{ks}
\end{gather}
The largest unstable mode may also be predicted, again assuming $k \gg 1$, by setting $\sigma \approx -k^4+(\Omega/2-F)k^2=0$, yielding $k_{max}\approx \sqrt{\Omega-2F}/\sqrt{2}=\sqrt{2}\,k^*$. These approximations are shown as dashed lines in Fig.~\ref{Figure3}d. Both overestimate their true values in the present study but capture the numerical results qualitatively. The reason for this discrepancy is due to the range of relevant wavenumbers involved in the elastocapillary instability. Destabilization of the bending modes occurs at small or intermediate values of the wavenumber $k$; the non-local corrections to the Euler-buckling problem ($\Omega g(k)$ terms in Eq.~\ref{dispersion}) thus cannot be overlooked and strongly contribute to the dynamic response of floating microfilaments.

In summary, we have shown in this section through a linear stability analysis that a self-attracting filament displays a novel mechanical instability. We have explored the structure of this instability, and its dependence on both the strength of the self-attraction and the stabilizing external force.

\section{Beyond linear-stability: self-folding of a floating filament}

The linear stability analysis of the previous section yields some insight into how folding will begin. In this section we investigate long-term, large-amplitude folding behavior. The path that a filament takes towards any final configuration depends on the initial conditions, the material parameters of the system (specified completely by $\Omega$), and the external tension on the filament ($F$). In this section we  compute the dynamics from the equations of motion described in \S III using the numerical method described in Appendix B. 

\begin{figure*}[t!]
\includegraphics[width=.99\textwidth]{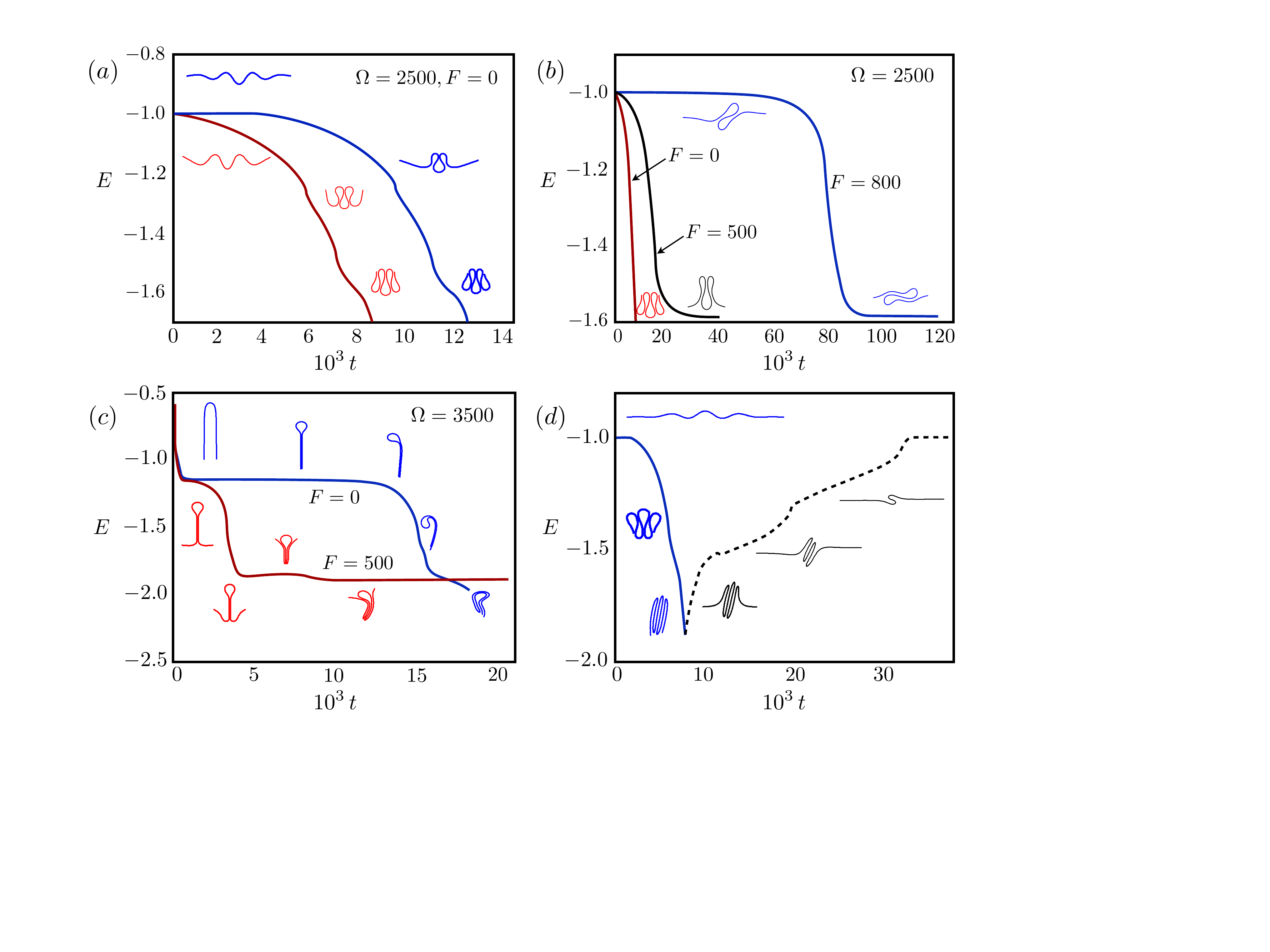}
\caption{Long-time dynamics of filament shape. 
(a): With $\Omega=2500$ and $F=0$, two initial perturbations  with different initial wavenumbers ($k=\pi$, top and  $k=7\pi$, bottom)  fold into similar structures but at different times. 
(b): Adding an external force, $F$, modifies the folding shape of the filament; with $\Omega=2500$ and $F=500$, the final folded shape has fewer loops than in the $F=0$ case due to the straightening influence of the added tension. Meanwhile, for $F=800$ a secondary instability develops along the axial direction, leading to a final left-right asymmetric shape. 
(c): With $\Omega=3500$, a highly deformed initial hairpin configuration undergoes a cascade of self-folding events. 
(d): Folding and unfolding of a floating filament is shown for $\Omega=3500$; the filament self-folds with $F=0$ until $t=9\times10^{-3}$ at which time a pulling force $F=3500$ is introduced and the filament  unzips, albeit through a different path through shape space.}
\label{Figure5}
\end{figure*}

For large values of the interaction parameter the capillary attraction overcomes the relatively weak bending resistance, and the filament displays a tendency to ``accordion": it collapses with a short wavelength relative to the filament length. The dynamics of such filament self-folding are shown in Fig.~\ref{Figure5}a, where we plot the value of the total energy $E$ (Eq.~\ref{Energy}) as a function of time, fixing $\Omega=2500$ and $F=0$ for two different initial filament shapes.
The topmost curve (blue online) shows the path taken of a filament that is seeded with an initial transverse position $y(s,0)=0.05 \sin(\pi s)$ as it travels down to a state of lower energy. At short times the shape is predominately composed of a number of anti-nodes corresponding to the fastest growing wave number,  predicted by maximizing Eq.~\ref{dispersion} to be  $k^*=7.4\pi$. These anti-nodes grow to form loops, and eventually the filament approaches a regime of near-contact. In this vicinity there is a slight discontinuity in the slope of the energy, as a result of introducing near-field repulsive physics required to regularize the diverging capillary interactions (see Appendix A).  The lower curve in Fig.~\ref{Figure5}a (red online) shows the path taken by an identical filament seeded with a different initial condition, $y(s,0)=0.05 \sin(7\pi s)$. Here the filament eventually achieves the same conformation as in the previously considered case, dictated by the fastest growing linear mode. However, due to its initial shape being closer to the fastest growing linearly unstable mode at the outset,  this filament begins its descent into lower energy conformations earlier. Interestingly, this result suggests that the elastocapillary buckling provides a robust means to achieve  complex foldings of microfilaments. Indeed, the gross features of the ultimate shapes can be anticipated from a simple linear stability analysis.

By introducing an external tension, $F>0$, the possibilities for folding become more versatile but more complex. In Fig.~\ref{Figure5}b we show three different filaments, all with initial positions given by $y(s,0)=0.05\sin (\pi s+0.01s)$ (a shape with a slight left-right asymmetry), but pulled upon with three different magnitudes of external forcing ($F=0$, 500, and 800). With the addition of this external tension the most unstable mode structure is changed, and a different final state is selected. With no external tension $(F=0)$, the filament accordions tightly as in Fig.~\ref{Figure5}a. In the second case, with $F=500$, the fastest growing unstable wavenumber is shifted to a smaller value, as predicted by the linear theory, leading to a smaller number of loops in an equilibrium conformation. Finally, with $F=800$ we observe a secondary instability growing along the axial direction, perpendicular to the transverse instability at the center of the previous section. 

In fact, when a filament is sufficiently long relative to the structures generated by the initial instability, it can undergo a cascade of instabilities. This is illustrated in Fig.~\ref{Figure5}c, where we seed the filament initially with the shape of a hairpin. In the absence of external forcing, the hairpin shape is seen to first ``zip up" into a racquet shape, reminiscent of the ones observed in wetted fibers~\cite{ElastoCapReview}, and in semi-flexible polymer systems~\cite{MackintoshI,MackintoshII}. Over time  it displays a cascade of  buckling and folding. By pulling on the filament ends with an external force, $F=500$, similar multiply folded states are obtained, yet the final shape is markedly different from the one observed in the force-free case. 


Finally, in Fig.~\ref{Figure5}d we illustrate the unfolding of a self-folded structure. The filament is set initially with the shape $y(s,0)=0.05\sin(7\pi s)$, and we fix $\Omega=3500$ and $F=0$. The shape evolves in a similar fashion as in the previously described examples (blue solid line). At $t=9\times10^{-3}$, we introduce external forcing on the filament ends with magnitude $F=3500$ to unfold (black dashed line). We observe an unzipping of the folded structure, with jumps in the filament energy indicative of each loop being pulled apart in succession. Rather than a reversible shape change we clearly observe unfolding along a different path.

\section{Conclusion}

In this paper we have described the elastocapillary instability of flexible filaments confined to a fluid surface. Buckling and folding result from the interplay between self-attraction through long-range capillary forces and elastic bending rigidity. As a result of the competition between these mechanical forces, the instability only appears beyond a large, critical ratio of the capillary self-attraction to the filament's bending modulus. The most unstable wavenumbers are typically located at an intermediate range, thereby yielding the spontaneous formation of accordions folds. Due to the the long-range nature of the capillary interactions, both the wavelength and the growth rate of this instability differ from the conventional Euler-buckling results, which would be observed for short-range self-attraction.

After having analyzed the onset of filament shape instability through a linear theory, numerical simulation was used to show a number of possible conformational transitions for the self-folding filament. In particular, introducing an external pulling force at the filament ends changes the unstable linear modes, and modifies significantly the long-time  behavior. For a filament sufficiently long relative to the shapes driven by self-attraction there can be a cascade of folding instabilities, each acting perpendicularly to the previous fold. While the linear analysis is only valid for short times and nearly straight filaments, our simulations demonstrate a wide variety of accessible conformations, from rackets and sewing needles to accordion shapes. 

The numerical method described here could be augmented to study many related problems. Future work might include computing the relative energy scales for each equilibrium shape; with sufficient statistics, the ensemble of different shapes that could be expected should our system be coupled to a thermal bath might be elucidated. The deterministic simulations could also be used to determine the statistical significance of certain shapes in crumpled thin sheets or fibers \cite{CrumplingStats,Crumplestats2}. 


Our original experimental motivation was provided by buoyant particle interactions, where the dominant competition is between elasticity and the density of the particles. In which practical experimental setup could the buckling instability be observed? We recall that the critical elastocapillary number is given by $\Omega=\gamma L^3/B$. The instability is expected to be observable for filaments both  soft (small value of $B$) and dense (large value of $\gamma$). Consider a filament composed  of microns-sized spherical colloids  linked by double-stranded DNAs as in Ref.~\cite{Dreyfus2005}. From capillary interactions we expect $\gamma\sim F^2/\sigma$ where $F$ is the force per unit length acting on the filament and $\sigma$ the surface tension of the fluid interface. From gravity we get $F\sim \Delta \rho a^2 g$ and thus  $\gamma\sim \Delta \rho^2 a^4 g^2/\sigma$. For colloids of radius $a=10\,\mu$m and $\Delta \rho=10^3$\,kg/m$^3$ and an interface with $\sigma\sim 10^{-2}$ N/m we would get 
$\gamma\sim10^{-10}$ N/m. Using filaments of length $L\sim 1$\,mm, right below the capillary length, and with a DNA linker flexibility of $B\sim 10^{-22}$ J/m \cite{Dreyfus2005}, we would obtain an elastocapillary number or order $\Omega\sim 10^3$, and thus the instability could be observable. Obviously, beyond the case of gravitational forcing, any external force acting on the filament can serve to deform the interface and drive similar elastocapillary self-folding. Electric or magnetic fields acting on self-assembled filaments of  metallic  nano-particles or super-paramagnetic nano filaments could be two experimental examples.

\section{Acknowledgements}

We acknowledge funding from the NSF (grant CBET-0746285).

\appendix

\section{Numerical solution}

The numerical approach taken in this manuscript is a variation on that described by Tornberg \& Shelley \cite{Tornberg2004}. The complications present here are the stiffness introduced by the elastic restorative term $\propto \b{x}_{ssss}$, and the nonlocal integral operator $\b{K}[\b{x}(s')]$. The approach will be to treat the stiffest terms implicitly in time, while treating the remaining terms explicitly for a fast but stable scheme. 

Time is discretized uniformly, with the $n^{th}$ timestep denoted by $t_n$, and writing (for instance) the tension at the $n^{th}$ timestep in shorthand as $T(s,t_n)=T^n(s)$. At each time $t_n$ the tension is first solved via
\begin{gather}
\left(\partial_{ss}-|\b{x}^n_{ss}|^2\right)T^n(s)=-A\left[4\b{x}^n_{ss}\cdot \b{x}^n_{ssss}+3|\b{x}^n_{sss}|^2\right]-c_s(s)-\\\nonumber\b{x}^n_s\cdot \frac{\partial}{\partial s} \b{K}[\b{x}^n(s')](s)+\beta(1-\b{x}_s\cdot\b{x}_s).
\end{gather}
We have included a correction term to remove numerical errors in the determination of the arc-length, again following \cite{Tornberg2004}. To solve this equation we define a uniform grid $s = i/M$, for $i=0,1,...,M$. Spatial derivatives computed using second-order accurate finite difference formulae. Since the position $\b{x}^n$ at time $t_n$ is known, the equation above is inverted to recover $T_n(s)$ without difficulty.

Solving the filament position equation is the next step, where we update via
\begin{gather}
\frac{1}{2\Delta t}\left(3 \b{x}^{n+1}-4\b{x}^n+\b{x}^{n-1}\right)=-\b{x}_{ssss}^{n+1}+\Omega c(s)(2\b{x}_s^n-\b{x}_s^{n-1})+\\\nonumber\Omega\b{K}[2\b{x}^n(s')-\b{x}^{n-1}(s')](s)+\partial_s\left(2T^n \b{x}^n_s-T^{n-1}\b{x}_s^{n-1}\right).
\end{gather} 
For the first time step the above is replaced by a first order Euler step. In this case, we need only build the following matrix operator and to invert it once: $\b{M}=3/(2\Delta t)\b{I}+ \b{D}_s^4$ (along with boundary conditions). The inverted operator may be applied to known quantities at every subsequent time step, providing for a very efficient numerical scheme. We have subtracted off the singular part inside the integrand of the interaction forcing so that the remainder of the expression is finite, as described in the main text.

\begin{figure}[t]
\begin{center}
\includegraphics[width=.27\textwidth]{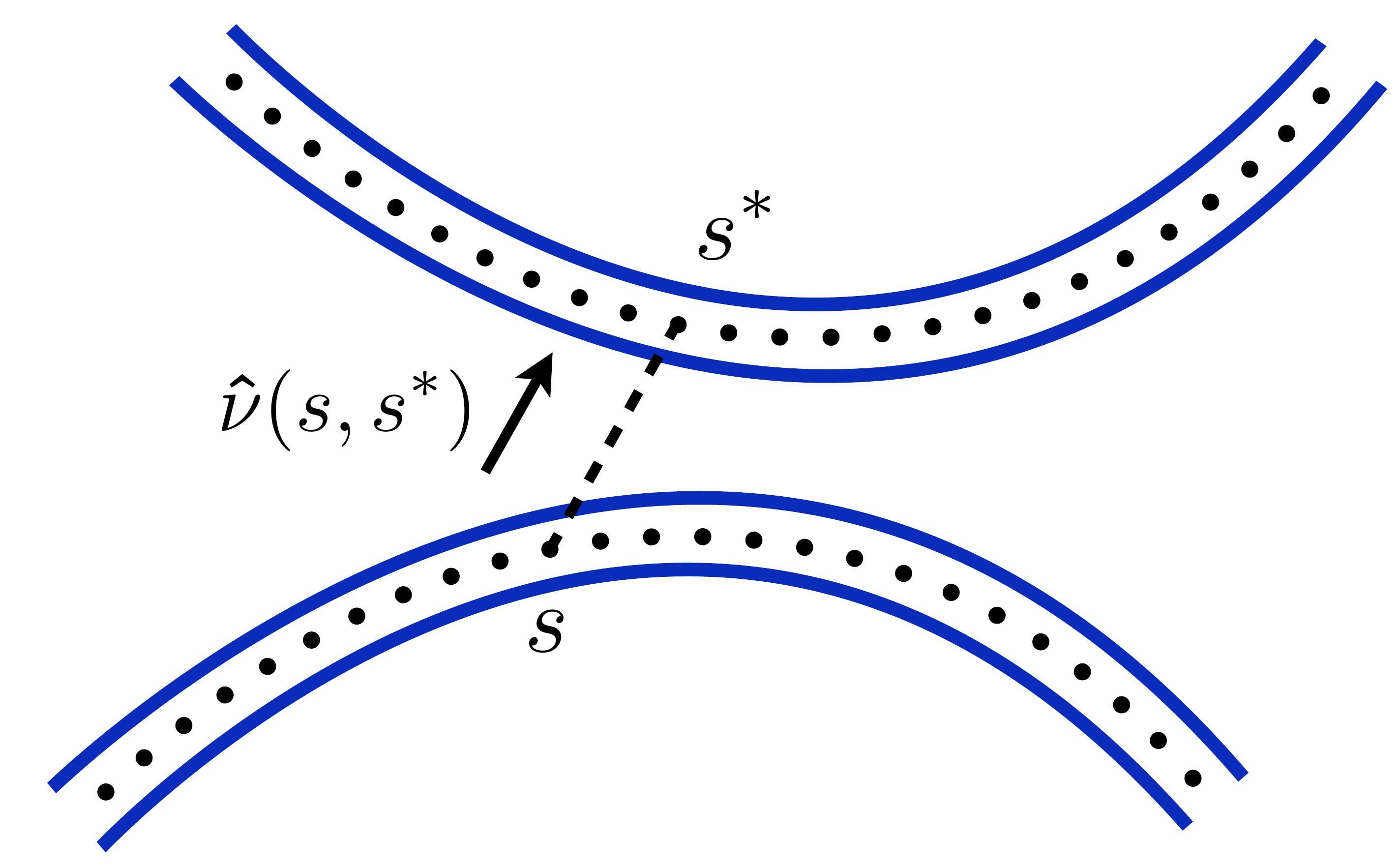}
\caption{A near-contact situation. $s^*$ denotes the point on the filament closest to the station $s$.}
\label{Figure6}
\end{center}
\end{figure}

A challenging numerical problem is introduced when the filament comes into near contact with itself. Or worse, when the filament exhibits a large number of loops so that there are many near-contact interactions. The situation is illustrated in Fig.~\ref{Figure6}, where two different parts of the filament are shown nearing each other; the centerline is denoted by dotted lines, and the blue solid lines indicate the thickness of the filament.

Our intention is to prevent filament crossing, but to allow for tangential motions in the near-contact state. For each station in arc-length on the filament $s$, we first determine the closest point on the filament $s^*$ that is at least a distance $4\e$ from $s$ (so as to neglect neighboring points on the filament),
\begin{gather}
s^*(s)=\min_{s'} \left\{ |\b{x}(s')-\b{x}(s)| \,\,\, , |s'-s|>4\e \right\}.
\end{gather}

A local coordinate system is then defined by $\b{\hat{\nu}}(s,s^*)=(\b{x}(s^*)-\b{x}(s))/|\b{x}(s^*)-\b{x}(s)|$ and $\b{\hat{\eta}}=\b{\hat{\nu}}^\perp$. The forces in the filament equation are then decomposed locally into their $\b{\hat{\nu}}$ and $\b{\hat{\eta}}$ components. The tangential component of the filament force balance is left unchanged, but the normal component is adjusted as follows
\begin{gather}
\b{\hat{\nu}}(s,s^*)\cdot \left(\b{x}_t+\b{x}_{ssss}\right)=\\
\nonumber\min\left\{\b{\hat{\nu}}(s,s^*)\cdot \b{G},\b{\hat{\nu}}(s,s^*)\cdot \b{G}\Big[1-\left(\frac{2\e}{|\b{x}(s)-\b{x}(s^*)|}\right)^p\Big]\right\}
\end{gather}
where $\b{G}$ denotes the right hand side of Eq.~\ref{filpos} (minus the stiffest term, $\b{x}_{ssss}$), and $p>0$ (we use $p=12$ for all the results presented here). The multiplication factor is such that there is no normal component of force (outside of the elastic part) when the filament centerline is $2\e$ away from the closest point on the other part of the filament centerline. The minimum of the two quantities is chosen so that if the fluid and tension forces wish to separate the filaments from each other they can, but if the fluid and tension forces are pushing the filaments to an overlapped state the sign of this force is reversed. This treatment of the near-contact physics appears to allow the filament to fold up tightly without numerical difficulties, without significantly impacting the underlying physics of capillary attraction. When the filament comes into near-contact, the timestep size is decreased and the spatial resolution is increased adaptively. 

For the linear stability analysis we set $M=2000$, $\Delta t=10^{-9}$, $\beta=5*10^6$. For the long-time dynamics and equilibrium shapes it was sufficient to set $M=600$, $\Delta t=10^{-8}$, $\beta=5*10^6$, and $p=12$, with adaptive refinement of the grid and decrease in timestep size as necessary when the filament approached itself as described.

\section{Nonlocal integration}

We begin by showing, for $k\gg 1$ and $s$ sufficiently far removed from the filament endpoints, that
\begin{gather}
K[\tilde{y}(s)]= \int_{0}^{1}\Big\{\frac{\tilde{y}(s')-\tilde{y}(s)}{(s'-s)^2}-\frac{\tilde{y}_s(s)}{s'-s}\Big\}\,ds'\approx -\pi k \tilde{y}(s),\label{B1}
\end{gather}
for $\tilde{y}(s)\propto \sin(ks)$. For $k \gg 1 $, we are led to write the integration over a new inner variable $s' = s + \xi/k$, so that 
\begin{gather}
K[\tilde{y}(s)] = k \int_{-k s}^{k(1-s)}\Big\{\frac{\sin(k s')-\sin(k s)}{\xi^2}-\frac{\cos(ks)}{\xi}\Big\}\,d\xi,
\end{gather}
and for $s$ well distanced from both endpoints, we may approximate the above as
\begin{gather}
K[\tilde{y}(s)] \approx k \int_{-\infty}^{\infty}\Big\{\frac{\sin(k s+\xi)-\sin(k s)}{\xi^2}-\frac{\cos(ks)}{\xi}\Big\}\,d\xi  \nonumber\\
=k \sin(ks) \int_{-\infty}^{\infty}\frac{\left(\cos(\xi)-1\right)}{\xi^2}\,d\xi+k \cos(ks) \int_{-\infty}^{\infty}\frac{\sin(\xi)-\xi}{\xi^2}\,d\xi \nonumber\\
=-\pi k \sin(ks) = -\pi k \tilde{y}(s).
\end{gather}
Though we do not reproduce it here, a complicated analytical expression for $K[\tilde{y}(s)]$ in Eq.~\ref{B1} may in fact be derived for arbitrary wavenumber $k$. While it is not generally the case that $K[\tilde{y}(s)]$ is diagonalized using a trigonometric basis (as it is in the $k \gg 1$ approximation above), we still note that 
\begin{gather}
\int_{0}^1 e^{-i k s} K[e^{i k s'}]\,ds = g(k),\\
g(k)=2[1-\cos(k)+\gamma + \log(k)-k \,S_i(k)-C_i(k)],
\end{gather}
with the sin and cosine integrals defined in Eq.~\ref{sici}. For $k \ll 1$ we have $g(k) \sim -k^2/2$, while for $k\gg 1$ we have $g(k) \sim -\pi k+2(\log(k)+\gamma+1)$, with $\gamma$ the Euler-Mascheroni constant. This latter approximation of $g(k)$ is accurate to within $2\%$ for wavenumbers as small as $k=\pi$.

\bibliography{SelfFold.bib}

\end{document}